# Repetition rate stabilization of an optical frequency comb based on solid-state laser technology with an intra-cavity electro-optic modulator

**NICOLAS TORCHEBOEUF, GILLES BUCHS, STEFAN KUNDERMANN, ERWIN PORTUONDO-CAMPA, JONATHAN BENNÈS AND STEVE LECOMTE\***

*Centre Suisse d'Electronique et de Microtechnique (CSEM), Jaquet-Droz 1, 2000 Neuchâtel, Switzerland*
*\*steve.lecomte@csem.ch*

**Abstract:** The repetition rate stabilization of an optical frequency comb based on diode-pumped solid-state laser technology is demonstrated using an intra-cavity electro-optic modulator. The large feedback bandwidth of such modulators allows disciplining the comb repetition rate on a cavity-stabilized continuous-wave laser with a locking bandwidth up to 700 kHz. This surpasses what can be achieved with any other type of actuator reported so far. An in-loop integrated phase noise of 133 mrad has been measured and the PM-to-AM coupling of the electro-optic modulator has been investigated as well.

### References and links

1. N. R. Newbury, "Searching for applications with a fine-tooth comb," Nat. Photonics **5**(4), 186–188 (2011).
2. S. A. Diddams, "The evolving optical frequency comb," J. Opt. Soc. Am. B **27**(11), 51–62 (2010).
3. J. Lee, Y.-J. Kim, K. Lee, S. Lee, and S.-W. Kim, "Time-of-flight measurement with femtosecond light pulses," Nat. Photonics **4**(10), 716–720 (2010).
4. I. Coddington, W. C. Swann, L. Nenadovic, and N. R. Newbury, "Rapid and precise absolute distance measurements at long range," Nat. Photonics **3**(6), 351–356 (2009).
5. E. Baumann, F. R. Giorgetta, J.-D. Deschênes, W. C. Swann, I. Coddington, and N. R. Newbury, "Comb-calibrated laser ranging for three-dimensional surface profiling with micrometer-level precision at a distance,"Opt. Express **22**(21), 24914-24928 (2014).
6. F. R. Giorgetta, W. C. Swann, L. C. Sinclair, E. Baumann, I. Coddington, and N. R. Newbury, "Optical two-way time and frequency transfer over free space," Nat. Photonics **7**(6), 434–438 (2013).
7. G. Marra, H. S. Margolis, and D. J. Richardson, "Dissemination of an optical frequency comb over fiber with 3 × 10^−18 fractional accuracy," Opt. Express **20**(2), 1843–1845 (2012).
8. T. M. Fortier, M. S. Kirchner, F. Quinlan, J. Taylor, J. C. Bergquist, T. Rosenband, N. Lemke, A. Ludlow, Y. Jiang, C. W. Oates, and S. A. Diddams, "Generation of Ultrastable Microwaves via Optical Frequency Division," Nat. Photonics **5**(7), 425–429 (2011).
9. F. Quinlan, F. N. Baynes, T. M. Fortier, Q. Zhou, A. Cross, J. C. Campbell, and S. A. Diddams, "Optical amplification and pulse interleaving for low-noise photonic microwave generation," Opt. Lett. **39**(6), 1581–1584 (2014).
10. E. Portuondo-Campa, G. Buchs, S. Kundermann, L. Balet, and S. Lecomte, "Ultra-low phase-noise microwave generation using a diode-pumped solid-state laser based frequency comb and a polarization-maintaining pulse interleaver," Opt. Express **23**(25), 32441–32451 (2015).
11. X. Xie, R. Bouchand, D. Nicolodi, M. Giunta, W. Hänsel, M. Lezius, A. Joshi, S. Datta, C. Alexandre, M. Lours, P.-A. Tremblin, G. Santarelli, R. Holzwarth, and Y. Le Coq, "Photonic microwave signals with zeptosecond-level absolute timing noise", Nat. Photonics, advance online publication 2016 (DOI 10.1038/nphoton.2016.215).
12. S. A. Diddams, L. Hollberg, and V. Mbele, "Molecular fingerprinting with the resolved modes of a femtosecond laser frequency comb," Nature **445**, 627–630 (2007).
13. B. J. Bjork, T. Q. Bui, O. H. Heckl, P. B. Changala, B. Spaun, P. Heu, D. Follman, C. Deutsch, G. D. Cole, M. Aspelmeyer, M. Okumura, and J. Ye, "Direct Frequency Comb Measurement of OD + CO → DOCO Kinetics," Science **354**(6311), 444–448 (2016).
14. T. J. Kippenberg, R. Holzwarth, and S. A. Diddams, "Microresonator-based optical frequency combs," Science **332**(6029), 555–559 (2011).

15. T. D. Shoji, W. Xie, K. L. Silverman, A. Feldman, T. Harvey, R. P. Mirin, and T. R. Schibli, "Ultra-low-noise monolithic mode-locked solid-state laser," Optica **3**(9), 995–998 (2016).
16. S. A. Meyer, T. M. Fortier, S. Lecomte, and S. A. Diddams, "A frequency-stabilized Yb:KYW femtosecond laser frequency comb and its application to low-phase-noise microwave generation," Appl. Phys. B **112**(4), 565–570 (2013).
17. E. Portuondo-Campa, R. Paschotta, and S. Lecomte, "Sub-100 attosecond timing jitter from low-noise passively mode-locked solid-state laser at telecom wavelength," Opt. Lett. **38**(15), 2650–2653 (2013).
18. A. Schlatter, B. Rudin, S. C. Zeller, R. Paschotta, G. J. Spühler, L. Krainer, N. Haverkamp, H. R. Telle, and U. Keller, "Nearly quantum-noise-limited timing jitter from miniature Er:Yb:glass lasers," Opt. Lett. **30**(12), 1536–1538 (2005).
19. D. Hou, C. Lee, Z. Yang, and T. R. Schibli, "Timing jitter characterization of mode-locked lasers with <1 zs/√Hz resolution using a simple optical heterodyne technique," Opt. Lett. **40**(13), 2985–2988 (2015).
20. G. Buchs, S. Kundermann, E. Portuondo-Campa, and S. Lecomte, "Radiation hard mode-locked laser suitable as a spaceborne frequency comb," Opt. Express **23**(8), 9890–9900 (2015).
21. S. Schilt and T. Südmeyer, "Carrier-Envelope Offset Stabilized Ultrafast Diode-Pumped Solid-State Lasers," Appl. Sci. **5**(4), 787–816 (2015).
22. H. R. Telle, G. Steinmeyer, A. E. Dunlop, J. Stenger, D. H. Sutter, and U. Keller, "Carrier-envelope offset phase control: A novel concept for absolute optical frequency measurement and ultrashort pulse generation," Appl. Phys. B **69**, 327–332 (1999).
23. J. Rauschenberger, T. M. Fortier, D. J. Jones, J. Ye, and S. T. Cundiff, "Control of the frequency comb from a mode-locked Erbium-doped fiber laser," Opt. Express **10**(24), 1404–1410 (2002).
24. C.-C. Lee, C. Mohr, J. Bethge, S. Suzuki, M. E. Fermann, I. Hartl, and T. R. Schibli, "Frequency comb stabilization with bandwidth beyond the limit of gain lifetime by an intracavity graphene electro-optic modulator," Opt. Lett. **37**(15), 3084–3086 (2012).
25. M. Hoffmann, S. Schilt, and T. Südmeyer, "CEO stabilization of a femtosecond laser using a SESAM as fast opto-optical modulator," Opt. Express **21**(24), 30054–30064 (2013).
26. L. Karlen, G. Buchs, E. Portuondo-Campa, and S. Lecomte, "Efficient carrier-envelope offset frequency stabilization through gain modulation via stimulated emission," Opt. Lett. **41**(2), 376–379 (2016).
27. D. D. Hudson, K. W. Holman, R. J. Jones, S. T. Cundiff, J. Ye, and D. J. Jones, "Mode-locked fiber laser frequency-controlled with an intracavity electro-optic modulator," Opt. Lett. **30**(21), 2948–2950 (2005).
28. E. Baumann, F. R. Giorgetta, J. W. Nicholson, W. C. Swann, I. Coddington, and N. R. Newbury, "High-performance, vibration-immune, fiber-laser frequency comb," Opt. Lett. **34**(5), 638–640 (2009).
29. Y. Nakajima, H. Inaba, K. Hosaka, K. Minoshima, A. Onae, M. Yasuda, T. Kohno, S. Kawato, T. Kobayashi, T. Katsuyama, and F.-L. Hong, "A multi-branch, fiber-based frequency comb with millihertz-level relative linewidths using an intra-cavity electro-optic modulator," Opt. Express **18**(2), 1667–1676 (2010).
30. K. Iwakuni, H. Inaba, Y. Nakajima, T. Kobayashi, K. Hosaka, A. Onae, and F.-L. Hong, "Narrow linewidth comb realized with a mode-locked fiber laser using an intra-cavity waveguide electro-optic modulator for high-speed control," Opt. Express **20**(13), 13769–13776 (2012).
31. W. Zhang, M. Lours, M. Fischer, R. Holzwarth, G. Santarelli, and Y. Le Coq, "Characterizing a fiber-based frequency comb with electro-optic modulator," IEEE Trans. Ultrason. Ferroelectr. Freq. Control **59**(3), 432–438 (2012).
32. T. C. Briles, D. C. Yost, A. Cingöz, J. Ye, and T. R. Schibli, "Simple piezoelectric-actuated mirror with 180 kHz servo bandwidth," Opt. Express **18**(10), 9739–9746 (2010).
33. N. Kuse, C.-C. Lee, J. Jiang, C. Mohr, T. R. Schibli, and M. E. Fermann, "Ultra-low noise all polarization-maintaining Er fiber-based optical frequency combs facilitated with a graphene modulator," Opt. Express **23**(19), 24342–24350 (2015).
34. N. Kuse, J. Jiang, C.-C. Lee, T. R. Schibli, and M. E. Fermann, "All polarization-maintaining Er fiber-based optical frequency combs with nonlinear amplifying loop mirror," Opt. Express **24**(3), 3095–3102 (2016).
35. M. Giunta, W. Haensel, K. Beha, M. Fischer, M. Lezius, and R. Holzwarth, "Ultra Low Noise Er:fiber Frequency Comb Comparison," in *Conference on Lasers and Electro-Optics*, OSA Technical Digest (1996) (Optical Society of America, 2016), paper STh4H.1.

## 1. Introduction

Highly stable optical frequency combs (OFCs) have led to numerous breakthroughs in the field of precision optical metrology and measurements [1,2] such as ultra-precise timing [3] and distance measurements [4,5] time and frequency transfer [6,7] generation of ultra-low phase noise microwaves [8–11] as well as optical spectroscopy [12] and even time-resolved spectroscopy in chemical reactions [13]. All these systems are mainly relying on modelocked laser technology while microresonator-based combs are currently intensively investigated [14]. In particular, modelocked lasers are very often based on fiber lasers but it turns out that solid-state lasers such as diode-pumped solid-state lasers (DPSSL) have fundamentally better noise properties than fiber lasers. Indeed, solid-state lasers that can be built in a very compact way

[15] have shown properties which are similar to those of Ti:sapphire lasers, like low cavity losses, high power and short pulse duration [16], leading to excellent timing jitter and very low amplitude noise [17–19]. Furthermore, solid-state lasers are particularly suited for operation in challenging environments as it has been recently reported in the context of space applications [20].

Most advanced applications listed above require a fully stabilized comb, implying that both the repetition rate ($f_{rep}$) and the carrier-envelope-offset (CEO) frequency ($f_{CEO}$) must be stabilized [21]. Detection of $f_{CEO}$ is typically implemented with the self-referencing technique using an *f*-to-2*f* interferometer [22] where the phase error signal is fed back to the pump power of the laser [23]. To circumvent the limitations of this method on the achievable phase lock bandwidth imposed by the gain medium stimulated lifetime [24], fast actuation methods have been demonstrated. These are: (i) cavity loss modulation through intra cavity graphene electro-optic modulators in fiber lasers [24], and (ii) loss modulation of the semiconductor saturable-absorber mirror (SESAM) [25] and gain modulation via stimulated emission in Er/Yb glass DPSSLs [26]. These methods allow for $f_{CEO}$ feedback bandwidths at least ten times higher than for the traditional gain modulation via pump power.

$f_{rep}$ stabilization is usually implemented by acting on the length of the laser cavity via a mirror mounted on an intra-cavity piezo-electric transducer. A control signal can be obtained either from direct photodetection of the laser pulse train, or from an optical beat note resulting from mixing the OFC spectrum with a reference cavity-stabilized CW-laser [8]. Faster actuation and thus larger stabilization bandwidths can be implemented using intra-cavity electro-optic modulators (EOMs) [27–31]. To the best of our knowledge, this approach has been implemented exclusively in fiber oscillators and no such demonstration in a solid-state laser has been implemented so far. Here, the EOM-based repetition rate stabilization of a DPSSL is presented, leading to a locking bandwidth up to 700 kHz.

## 2. Experimental setup

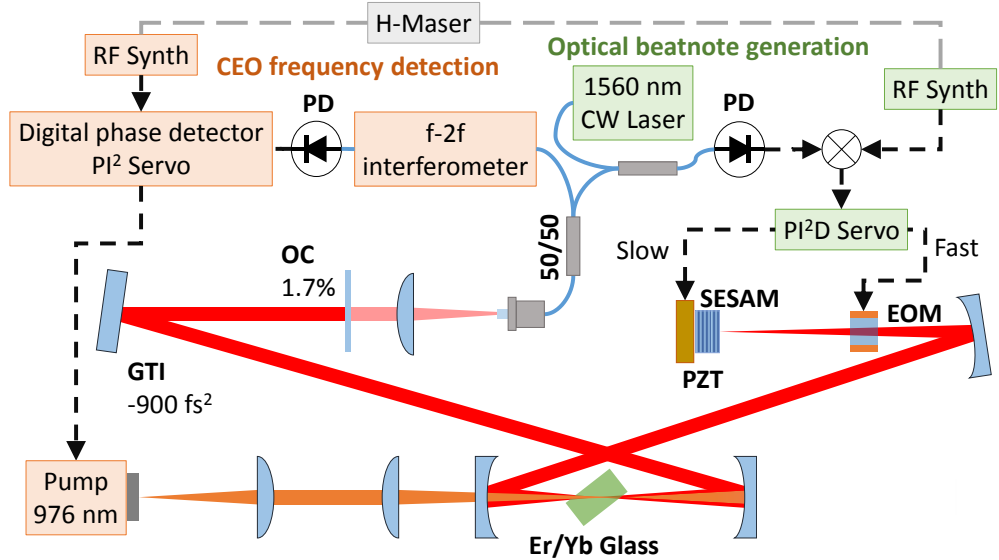

Fig. 1 . Schematic of the experimental setup with the two phase-locked loop (PLL) systems to respectively stabilize the carrier-envelope offset (CEO) frequency (red) and the repetition rate by generating an optical beat note (green). Er/Yb glass: codoped erbium-ytterbium glass, OC: output coupler, GTI: Gires-Tournois interferometer mirror, PD: photodetector, Semiconductor Saturable Absorber Mirror (SESAM) mounted on a piezoelectric transducer (PZT), EOM: electro-optic modulator.

The laser oscillator used in this work is an Er/Yb glass DPSSL which is passively modelocked with a semiconductor saturable absorber mirror (SESAM). A schematic of the full experimental setup is presented in

Fig. 1. The pump diode is a standard telecom fibered device emitting up to 600 mW at a wavelength of 976 nm. The laser architecture is very similar to the one reported in Ref. [26]. It emits transform-limited 220 fs soliton pulses at a center wavelength of 1546 nm with a repetition rate of 100.05 MHz. In order to fully stabilize the femtosecond laser, both $f_{CEO}$ and $f_{rep}$ have been locked.

More specifically, the frequency comb was self-referenced with an *f-2f* nonlinear interferometer. The carrier-envelope offset frequency $f_{CEO}$ signal was detected with a signal-to-noise ratio (SNR) of 50 dB in a 91 kHz resolution bandwidth (rbw). It was further phase-locked to a synthesizer referenced to a hydrogen maser. The correction signal was applied to the injection current of the laser pump diode. An integrated phase noise of 286 mrad [10 Hz; 1 MHz] has been typically achieved with a locking bandwidth of 45 kHz [10].

For $f_{rep}$ stabilization, a home-made cavity-stabilized continuous-wave (CW) laser with a narrow, 1-Hz-level linewidth at 1560 nm was used as reference. The comb light was filtered with a 1.2 nm bandpass filter centered at 1560 nm prior to mixing with the cavity-stabilized laser light. At the photodetector stage, a beat note generated by a single comb mode and the CW light was selected with an electrical bandpass filter, then amplified and sent to the input of a mixer that acts as phase detector. The second input of the phase detector was fed with a hydrogen maser referenced synthesizer signal. The generated error signal was then converted to a correction signal by means of a 10 MHz bandwidth analog PI$^2$D controller (Vescent Photonics, D2-125). This correction signal was further injected into a high-voltage source with a 1 MHz bandwidth for driving the intra-cavity EOM for fast actuation. The second output of the PI$^2$D controller was sent to a second high-voltage source for low-bandwidth large-range actuation of the laser cavity length via a piezoelectric transducer holding the SESAM.

The intra-cavity EOM consists in a 3mm-long z-cut lithium niobate crystal with anti-reflection coated facets. In order to avoid the formation of sub-cavities causing detrimental instabilities of the modelocked laser, the EOM has been inserted in a region where the laser mode is not collimated, in this case, between a curved mirror and the SESAM. Since the laser operates in the solitonic regime, the intra-cavity dispersion needs to be controlled and set to a negative value. The lithium niobate EOM naturally introduces about 660 fs$^2$ (110 fs$^2$/mm) of positive dispersion by round-trip. Thus, an additional cavity mirror with negative dispersion was inserted, counter balancing the positive dispersion of the EOM crystal. Taking into account all the added positive dispersion and knowing that the dispersion within the doped glass (-80 fs²) is insufficient to compensate it, it was necessary to introduce about -900 fs² by means of Gires-Tournois interferometer mirrors to get the laser into the soliton regime.

The EOM can act as a wave plate and thus generate polarization rotation in the cavity which leads to AM modulation via polarization selection of the gain medium oriented with Brewster angle. In order to minimize this effect, the orientation of the EOM was carefully adjusted. For this alignment, a modulation signal was introduced in the EOM. The oscillator output was photodetected and the AM peak corresponding to the EOM modulation frequency was monitored with an RF spectrum analyzer. Then, the EOM orientation was adjusted so as to minimize the AM peak.

### 3. Results

The insertion of the intra-cavity EOM causes additional cavity losses and thus impacts the laser performance in terms of output power and pulse duration. Without intra-cavity EOM, the oscillator produced 141 fs soliton pulses with 105 mW average power while, with the EOM, it delivered 220 fs soliton pulses with 55 mW average power. Also, the central wavelength of the oscillator was shifted by 9 nm, from 1555 to 1546nm after the addition of the EOM. Despite these changes of the oscillator parameters, the laser maintained a continuous modelocked regime for periods of days without intervention. The beat note for repetition rate locking, obtained from mixing the CW and the comb light, presented a SNR of 40 dB in 91 kHz rbw. It has to be noted that we usually measure a 10 dB higher SNR [10]. We attribute this reduction to the less optimal overlap of the frequency comb optical spectrum with the CW light wavelength.

The repetition rate was successfully stabilized by using the combination of the EOM for fast corrections and the piezo actuator for larger range and slower corrections. While no special effort to achieve long term stability was undertaken, the system remained locked for hours. The RF spectrum of the stabilized optical beat note is shown in Fig. 2. (a).

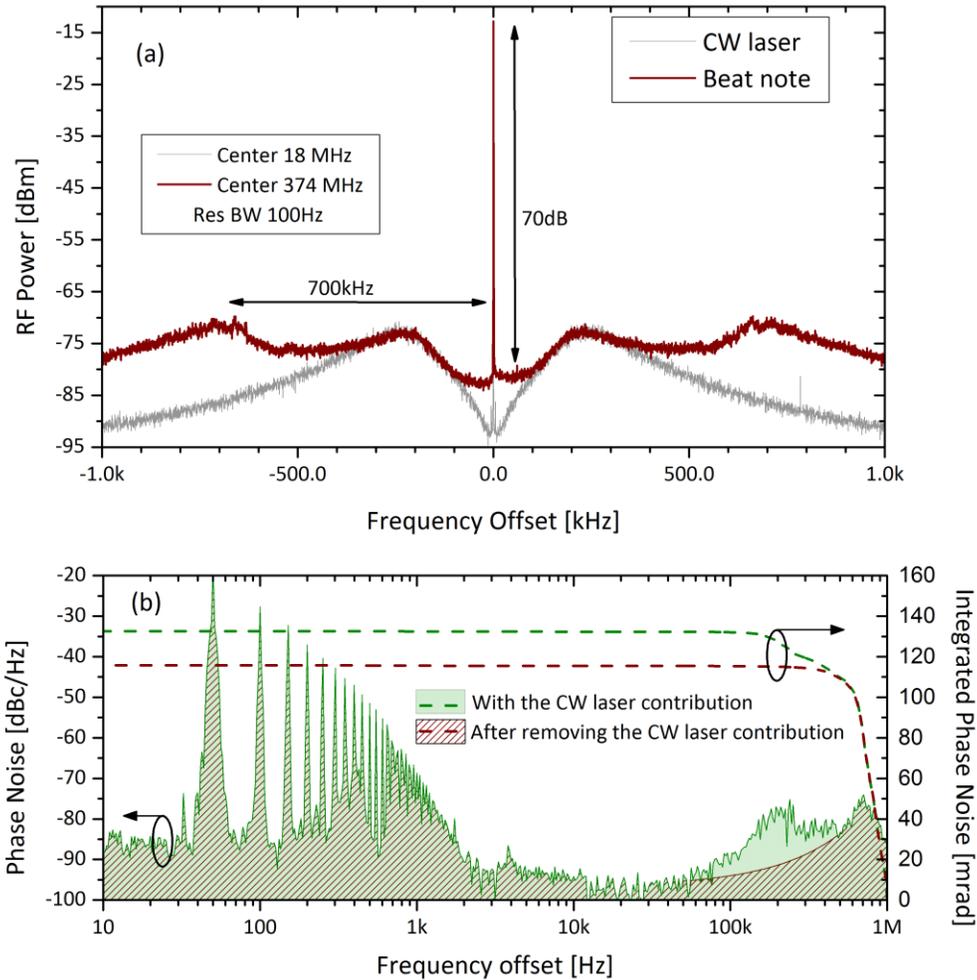

Fig. 2. (a) Red: In-loop RF spectrum of the stabilized optical beat note. Grey: In-loop RF spectrum of the cavity-stabilized CW laser around its phase modulation frequency of 18 MHz. Both spectra were measured with 100 Hz rbw. (b) Left: Phase noise spectrum. Right: Integrated phase noise. An integrated phase noise of 133 mrad has been measured in the range from 10 Hz to 1 MHz. After numerically removing the contribution from the CW laser the calculated integrated phase noise is 115 mrad [10 Hz; 1 MHz].

The repetition rate locking bandwidth limited by typical servo bumps is close to 700 kHz in Fig. 2(a). This is about a factor 4 larger than the best result reported for a solid-state laser using an optimized piezo actuator [32]. A more in-depth analysis can be made based on the measured phase noise spectrum presented in Fig. 2 (b) and which measured using an Agilent N9030A PXA Signal Analyzer.

One clearly recognizes the signature of the servo bump corresponding to the EOM-based lock at an offset frequency of about 700 kHz. At around 250 kHz, another servo bump feature is present and is attributed to the cavity-stabilized CW laser. In the range of 50 Hz to 2 kHz, the large number of spurs is attributed to the power line at 50 Hz and its harmonics. This technical noise source can in principle be eliminated with proper electro-magnetic shielding, which was not attempted in this work. An integrated phase noise of 133 mrad [10 Hz; 1 MHz] has been measured and, despite the limitations mentioned above (limited SNR, CW laser phase noise, power line induced spurs), this already represents a state-of-the-art result [33–35]. In Fig. 2 (b) the CW laser phase noise is numerically removed and the recalculated integrated phase noise amounts 115 mrad [10 Hz; 1 MHz].

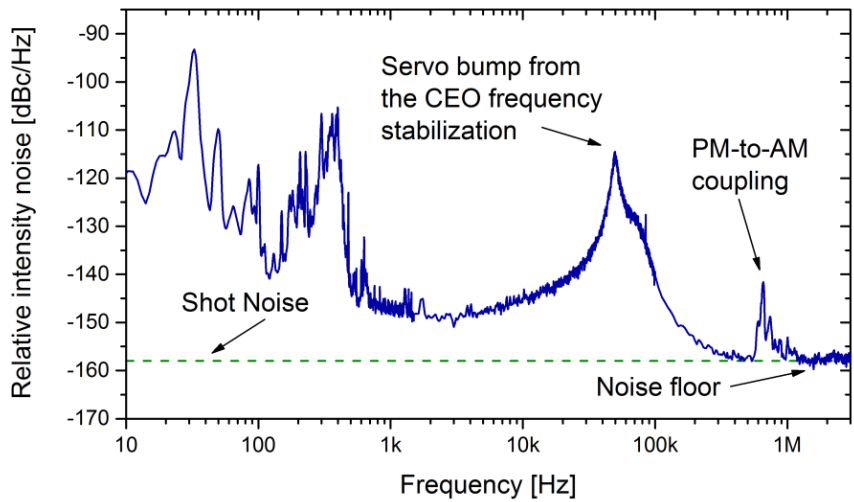

Fig. 3. Relative intensity noise of the fully stabilized laser. The servo bump from the stabilization of the carrier-envelope offset frequency appears at 45 kHz and a spur around 700 kHz can be interpreted as PM-to-AM coupling due to the EOM.

In the relative intensity noise (RIN) (Fig. 3.) a spur appears around 700 kHz, far from the carrier-envelope offset frequency servo bump which is at 45 kHz. It can be interpreted as PM-to-AM noise due to a slight misalignment of the EOM, generating a loss of power on the doped glass interface caused by the polarization change. A similar spur is also present on the in-loop carrier-envelope offset frequency spectrum.

### 4. Summary and outlook

The repetition rate stabilization of a diode-pumped solid-state laser based optical frequency comb by means of an intra-cavity EOM has been demonstrated. A locking bandwidth of about 700 kHz has been reached. This result is similar to typical bandwidth obtained with fiber lasers equipped with an intra-cavity EOM. It however sets a locking bandwidth record for a solid-state-based system [32]. An integrated phase noise of 133 mrad [10 Hz, 1 MHz] has been achieved and ways to improve this result have been identified and discussed. Such an intra-cavity electro-optical actuator also opens possibilities to control the repetition rate in recently reported monolithic solid-state lasers, attractive for their ultra-low noise performance [15].

**Funding**

European Commission (EC) (FP7-SPACE-2013-1-607087- PHASER); Canton de Neuchâtel.

**Acknowledgments**

The authors would like to thank L. Balet for support in data acquisition, T. J. Kippenberg from EPFL for the supervision of the master thesis of N. Torcheboeuf, S. Schilt from the University of Neuchâtel for the loan of the Vescent Photonics controller and F. Quinlan from NIST for useful discussions on intra-cavity EOMs. The financial support of the Canton of Neuchâtel is gratefully acknowledged.